\documentclass[amssymb, nobibnotes, aps, pra,floatfix,superscriptaddress,notitlepage,twocolumn]{revtex4-1}

\usepackage{graphicx} 
\usepackage{hyperref}
\usepackage{amsmath}
\makeatletter

\providecommand{\tabularnewline}{\\}

\makeatother

\begin{document}

\title{A New Light Higgs Boson and Short-Baseline Neutrino Anomalies}

\author{J. Asaadi}
\affiliation{Department of Physics, University of Texas at Arlington,
Arlington, TX 76019, USA}

\author{E. Church}
\affiliation{Pacific Northwest National Laboratory, PO Box 999,
Richland, WA 99352, USA}

\author{R. Guenette}
\affiliation{Department of Physics, Harvard University, Cambridge, MA 02138, USA}

\author{B.J.P. Jones}
\thanks{Corresponding author: \url{ben.jones@uta.edu}}
\affiliation{Department of Physics, University of Texas at Arlington,
Arlington, TX 76019, USA}

\author{A. M. Szelc}
\affiliation{School of Physics and Astronomy, University of Manchester, Manchester, M13 9PL, United Kingdom.}

\begin{abstract}
    The low-energy excesses observed by the MiniBooNE experiment have, to date, defied a convincing explanation under the standard model even with accommodation for non-zero neutrino mass.
    In this paper we explore a new oscillation mechanism to explain these anomalies, invoking a light neutrinophilic Higgs boson, conceived to induce a low Dirac neutrino mass in accord with experimental limits. Beam neutrinos forward-scattering off of a locally over-dense relic neutrino background give rise to a novel matter-effect with an energy-specific resonance. An enhanced oscillation around this resonance peak produces flavor transitions which are highly consistent with the MiniBooNE neutrino- and antineutrino-mode data sets. The model provides substantially improved $\chi^2$ values beyond either the no-oscillation hypothesis or the more commonly explored 3+1 sterile neutrino hypothesis.  This mechanism would introduce distinctive signatures at each baseline in the upcoming SBN program at Fermilab, presenting opportunities for further exploration.
\end{abstract}

\maketitle
\flushbottom

\section{Motivation \label{sec:Motivation}}

The nature of neutrino mass is widely recognized as one of the most important open theoretical and experimental questions in particle physics. In the Standard Model (SM) particles acquire mass via the Higgs mechanism, and the measured Yukawa couplings of the fermions span from $7\times10^{-1}$ (top quark) to $2\times10^{-6}$ (electron). Under the SM, neutrinos do not couple directly to the Higgs field and so are massless. However, the discovery of neutrino oscillation \cite{McDonald:2016ixn,Kajita:2016cak} has demonstrated that neutrinos have a tiny but non-zero mass. This mass has current upper limits around $m_\nu\lesssim 2.05\,\mathrm{eV}$ from direct searches \cite{Kraus:2004zw, Aseev:2011dq}. Cosmological observations of the cosmic microwave background and baryon acoustic oscillations \cite{Ade:2015xua} suggest the sum of neutrino masses is still smaller, $\sum_{i}m_{i}<0.23\,\mathrm{eV}$.  One neutrino mass state may still be massless $m_{\nu1}\geq0$, but oscillations \cite{Patrignani:2016xqp} place lower limits on the other two masses of $m_a>8.5\,meV$ and $m_b>50\,meV$ (the identification of subscript $a$ and $b$ with a conventionally numbered mass state depends on the ordering of the neutrino masses).  Such small values imply either extremely small Yukawa couplings to the Higgs or a new mechanism for mass generation. Several models have been proposed to try to explain the mass of neutrinos \cite{Chang:1985en,minkowski1977mu,gell1979ramond,yanagida1979proceedings,mohapatra1981neutrino}, many of them predicated on the seesaw mechanism and the generation of a Majorana neutrino mass via the Weinberg operator.

However, the possibility remains that neutrinos are Dirac particles, without a Majorana mass term. The challenge for such models is explaining why the Yukawa coupling to neutrinos are so much smaller than to the other particles \cite{Donoghue:2016tjk}. This would be particularly awkward, since the neutrinos are members of SU(2) doublets with the charged leptons, all of which
have much larger masses. Admitting Dirac neutrinos thus suggests nontrivial structure in the Higgs sector. This is often codified in two-Higgs-doublet models \cite{Branco:2011iw}. In particular, some authors have suggested that new light Higgs fields with a smaller
vacuum expectation value ({\em vev)} could generate the light neutrino masses, while the standard model Higgs generates the masses of the other fermions \cite{Wang:2006jy,Gabriel:2006ns,Baek:2016wml,Davidson:2009ha}. In this work we consider possible phenomenological implications of such a scheme.  We do not restrict our considerations to any specific embedding within the Standard Model, but rather treat it as a generic consequence of a wider class of models, some examples of which are listed in the above references.

We consider here a minimal model with a single new neutrinophilic Higgs boson that has a standard-model type potential:
\begin{equation}
\mathcal{L}_{\nu Higgs}=\epsilon^{2}\Phi^{\dagger}\Phi+\zeta\left(\Phi^{\dagger}\Phi\right)^{2} \mathrm{with} \quad\epsilon^{2}<0\quad\zeta>0,
\end{equation}
generating a new Higgs mass $m_{h}=\sqrt{2\zeta}v_{h}\quad>0$ and a {\em vev} $v_{h}=|\epsilon|/\sqrt{\zeta}\quad>0$. The masses of the neutrinos are then given by Yukawa couplings to new Higgs:
\begin{equation}
\mathcal{L}_{\nu mass}=\sum_{i}\left(g_{i}v_{h}\right)\bar{\nu}\nu.
\end{equation}
Both $m_h$ and $v_h$ values are independent free
parameters. To generate natural neutrino masses using Yukawa couplings of similar order to those in the standard Higgs sector the {\em vev} of
the new Higgs field should be in the range $10\,meV<v_{h}<100\,keV$.
It is worth noting that the ``natural'' range of Yukawas in this
new sector could be entirely unrelated to that in the known Higgs sector, so, as usual, the parameters required for naturalness cannot be stated unambiguously.  

The mass of the new boson is an independent and unconstrained parameter which is, in principle, experimentally measurable. If prejudices developed from the SM Higgs sector were directly applicable the Higgs self coupling $\epsilon$ would be $O(1)$, and the new Higgs mass
would be of the same order as the new {\em vev}. Again, this should only be taken as a rough guide. Finally, we note that the neutrinos in our model are strictly Dirac fermions with no Majorana term present, and we do not rely on a seesaw model to explain their mass scale. 

A particularly interesting phenomenological implication of this model is that exchange of this new boson with the cosmic neutrino background (CNB) can, under the circumstances discussed in this paper, generate a resonance observable in
neutrino oscillation experiments. In particular, we will show that for a Higgs mass of O(10keV) and a CNB over-density that is large but within experimental limits, this resonance may be observable in short-baseline neutrino oscillation experiments. In some cases, oscillation signatures
very similar to the MiniBooNE low-energy excess may be generated. 

\section{New Higgs exchange with the Cosmic Neutrino Background}

We consider a beam of relativistic neutrinos with a narrow momentum spread centered at $\left\{ E,p\right\} $ acting as test-particles. A neutrinophilic Higgs boson introduces a new force between the test particles
and other neutrinos but does not couple strongly to other Standard Model particles. These interactions, exhibited between the test particle and the pervasive bath of relic neutrinos produced in the hot big bang, may then generate observable phenomena.

The Higgs couplings to neutrinos are determined by the Yukawa parameters. Were Dirac neutrino masses generated by a standard model Higgs alone, this Yukawa coupling would be of order $10^{-15}<g<10^{-13}$, and so standard Higgs-to-neutrino interactions would be negligible. A larger contribution comes from $Z$-boson exchange with the relic neutrino background, which has been studied in \cite{Diaz:2015aua}, but its effects remain negligible for neutrino experiments.

In our model the {\em vev} of the Higgs is much smaller, so the Yukawa coupling used to generate the neutrino masses are large. Under such conditions, coherent forward scattering from the CNB via light Higgs exchange creates a refractive effect. Because this refractive effect depends on the mass composition of the beam, the phase of each mass eigenstate advances at different rates, thus increasing the frequency of neutrino oscillations.

\begin{figure}[t]

\centering
\includegraphics[width=0.9\columnwidth]{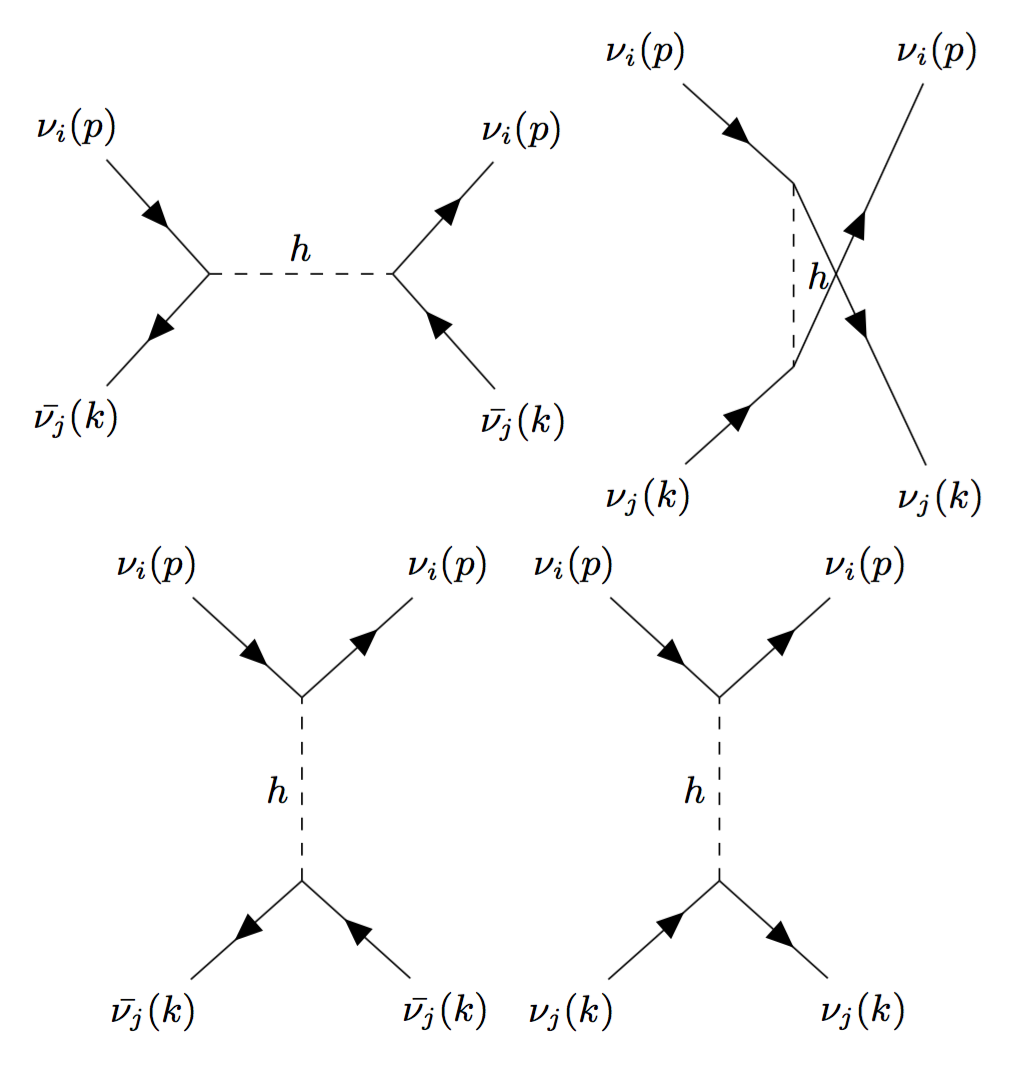}

\caption{Feynman diagrams for scattering of neutrinos off CNB neutrinos and anti-neutrinos. Top row: diagrams 1 and 2. Bottom row: diagrams 3 and 4. \label{fig:FeynamDiagrams}}
\end{figure}

The relevant Feynman diagrams for the interaction of beam neutrinos with CNB neutrinos are shown in Fig.~\ref{fig:FeynamDiagrams}. We label each diagram with an amplitude ${\cal M}_{ij}^{a}(p,k)$ where the
index $a$ specifies whether a relic $\nu$ or $\bar{\nu}$ is
involved, $p$ labels the four-momentum of the test neutrino from the beam, decomposed in terms of mass eigenstates $i$, and $k$ labels the four-momentum of the relic neutrino or antineutrino, decomposed in terms of mass eigenstates $j$. We describe the calculation for neutrinos, but exactly analogous effects are present for antineutrinos as test particles. 

To maintain coherence with the incident beam required for oscillation, we are limited to interactions where the incoming and outgoing momenta and masses of both the test particle and the relic neutrino are unchanged. Given that in our model the mass of the neutrinos is defined by diagonality of the Higgs coupling, it is immediately clear that the first two diagrams only give nonzero contributions when $i=j$

Applying the Feynman rules, allowing for a finite Higgs width, and
assuming the relic neutrinos to be unpolarized, we find
the spin-averaged matrix elements:
\begin{align}
i\langle{\cal M}_{ij}^{\bar{\nu},1}\rangle ={}&i\frac{g_{i}^{2}\left[k.p-m_{i}^{2}\right]}{\left(p+k\right)^{2}-m_{h}^{2}-im_{h}\Gamma_{h}}\delta_{ij} \label{eq:M1} \\
i\langle{\cal M}_{ij}^{\nu,2}\rangle={}&i\frac{g_{i}^{2}\left[k.p+m_{i}^{2}\right]}{\left(p-k\right)^{2}-m_{h}^{2}-im_{h}\Gamma_{h}}\delta_{ij} \label{eq:M2} \\
i\langle{\cal M}_{ij}^{\bar{\nu},3}\rangle={}&2i\frac{g_{i}^{2}g_{j}^{2}\nu_{h}^{2}}{m_{h}^{2}-im_{h}\Gamma_{h}} \label{eq:M3} \\
i\langle{\cal M}_{ij}^{\nu,4}\rangle={}&-2i\frac{g_{i}^{2}g_{j}^{2}\nu_{h}^{2}}{m_{h}^{2}-im_{h}\Gamma_{h}} \label{eq:M4}
\end{align}
To account for the complete refractive effect of the relic background, we must sum over the background momentum and flavor distribution. This distribution
is determined by a density function, which in the standard
cosmological model is the Fermi-Dirac function, redshifted from decoupling
time to a temperature $T\sim1.95\,\mathrm{K}$ in the present era \cite{GiuntiKim}.

In the absence of a chemical potential
driving an over-density of either neutrinos or anti-neutrinos in the
early universe, $\mu_{i}=0$ and the distributions for $\nu$ and $\bar{\nu}$ are equivalent. The number density $\langle n_{i}\rangle$ of all mass states are also expected to be equivalent,
with $\langle n_{i}\rangle=56\,\mathrm{cm^{-3}}$, for each flavor, for both neutrinos and anti-neutrinos.   

With equal number densities of relic neutrinos and anti-neutrinos, the contributions of Feynman diagrams 2 and 3 cancel in the forward scattering
amplitude. We assume this case to simplify our calculations.  In practice the effects of any asymmetry, if present, would be highly sub-leading relative to resonant effects in our final expressions.

Small deviations from the baseline model described above are expected, arising from the effects
of gravitational clustering for non-relativistic neutrinos \cite{Ringwald:2004np,deSalas:2017wtt}. More dramatic clustering effects can, however, arise in the presence of new forces between neutrinos, through the process of ``neutrino cloud'' formation \cite{Stephenson:1996qj}, with enhancements of up to 10$^{14}$ considered in Ref.~\cite{McKellar:2001hk}. Local over-densities of $10^{9}$ have been discussed in connection
with direct relic neutrino detection experiments \cite{Kaboth:2010kf}, 
and are still outside of experimental reach. 
Independently, local enhancements of $10^{6}$ have been suggested \cite{Faessler:2016tjf,Lazauskas:2007da} in the context of direct detection, in order to match local baryon over-density of the galaxy.  Finally, over-densities of $10^{13}$ have been invoked in order to explain the knee of the cosmic-ray spectrum \cite{Hwang:2005dq}.  In this work we consider the local relic density as a free parameter, and find large observable effects within the range of values previously considered by others.

At temperatures of 2K, the energies of the relic neutrinos are in the $0.1-1$ meV range. Thus all except perhaps the lightest are non-relativistic, and we can reasonably approximate them to be at rest.
The dominant effects in our model will be caused by scattering from the species with the largest Yukawa coupling, that is, the heaviest neutrino mass state. These are necessarily non-relativistic in the present cosmological era, so we substitute the four vector $k=(m,0)$ and replace the sum over the
Fermi-Dirac distribution with a simple multiplication by number density.
With all these considerations, the only important scattering amplitudes
arise from diagrams 1 and 2, each involving a relic neutrino or
anti-neutrino in the same mass eigenstate as the test particle. If we make the
assumption that the new boson is truly neutrinophilic, we may substitute for its width:
\begin{equation}
\Gamma=\frac{1}{2}m_{h}\sum_i g_{i}^{2},
\end{equation}
to yield real and imaginary parts of the scattering matrix shown in equations \ref{eq:ReT} and \ref{eq:ImT}:
\begin{widetext}
\begin{equation}
Re\langle M\rangle=\delta_{ij}\frac{g_{i}^{2}E}{2m_{i}^{2}}\left[\frac{E-E_{res}^i}{\left(E-E_{res}^i\right)^{2}+\left(\frac{1}{2}E_{res}^ig_{i}^{2}\right)^{2}}+\frac{-E-E_{res}^i}{\left(E+E_{res}^i\right)^{2}+\left(\frac{1}{2}E_{res}^ig_{i}^{2}\right)^{2}}\right]\label{eq:ReT}
\end{equation}
\begin{equation}
Im\langle M\rangle=-\delta_{ij}\frac{g_{i}^{4}E}{4m_{i}^{2}}\left[\frac{E_{res}^i}{\left(E-E_{res}^i\right)^{2}+\left(\frac{1}{2}E_{res}^ig_{i}^{2}\right)^{2}}+\frac{E_{res}^i}{\left(E+E_{res}^i\right)^{2}+\left(\frac{1}{2}E_{res}^ig_{i}^{2}\right)^{2}}\right].\label{eq:ImT}
\end{equation}
\end{widetext}
In the above equations we introduce an important phenomenological  parameter, the resonance energy $E_{res}^i=m_{h}^{2}/2m_{i}$. It is  instructive to consider a more limited model invoking the Zero Width Approximation (ZWA), which involves setting $\Gamma\rightarrow0$ in equations \ref{eq:M1} and \ref{eq:M2}.  In practice, this will be a good approximation when the Yukawa couplings are in the range $g\leq0.1$.  In this case, we find:
\begin{equation}
Re\langle M_{ZWA}\rangle=\delta_{ij}\frac{g_{i}^{2}}{4m_{i}}\left[\frac{-E_{res}^i}{E^{2}-E_{res}^{2}}\right]\label{eq:ReT-1}
\end{equation}
\begin{equation}
Im\langle M_{ZWA}\rangle=0.\label{eq:ImT-1}
\end{equation}
In both the ZWA and finite-width cases, a clear resonance is observed
at $E=E_{res}^i$ for each mass state, corresponding to production of new Higgs bosons at rest in the center of mass frame. Although this resonance only has a small effect in terms of real particle production or scattering, as implied by the O($g_i^4$) suppression of $Im[M]$ and as demanded by the optical theorem,
it makes a significant contribution to the real part of the forward-scattering amplitude, and thus contributes a large oscillation phase near the resonance.\newline

\section{Connection to neutrino refractive properties and oscillations}

The scattering matrix calculated above has both real and imaginary parts, which contribute to the refractive and absorptive behaviors of the neutrino beam, respectively. It will be most convenient in
what follows to work with the $T$ matrix normalized with single-particle wave functions, rather than the Lorentz-invariant $M$ matrix. These are related by $T=M/4Em_{i}$.

For relativistic forward scattering we can incorporate the effects of the amplitudes calculated above into the time-evolution of the neutrino wave-function, as:
\begin{equation}
\psi'=u_{0}(p)Exp\left[i\left(p+in_{i}T_{i}\right)x\right].
\end{equation}
It is easy to verify that neutrino refractive properties reproducing the standard MSW matter potential are recovered if $T$ is chosen to be that of the weak interactions. In our case, the matrix elements of interest are given instead by eqs.~\ref{eq:M1} and \ref{eq:M2}.

Following the standard derivation of the neutrino oscillation formula, we find the probability for conversion from a neutrino flavor $\alpha$ to neutrino flavor $\beta$ is given by eq.~\ref{eq:OscProb} :

\begin{widetext}
\begin{equation}
P_{\alpha\beta}=\delta_{\alpha\beta}-4\sum_{i>j}Re(U_{\alpha i}^{*}U_{\beta i}U_{\alpha j}U_{\beta j}^{*})\mathrm{sin}^{2}\left[\frac{\Delta_{ij}}{2}L\right]+2\sum_{i>j}Im(U_{\alpha i}^{*}U_{\beta i}U_{\alpha j}U_{\beta j}^{*})\mathrm{sin}^{2}\left[\frac{\Delta_{ij}}{4}L\right] \label{eq:OscProb}
\end{equation}
\end{widetext}

with $\Delta_{ij}=\left(p_{i}-p_{j}\right)+nRe\left(T_{i}-T_{j}\right)
$. Noting that, as usual:
\begin{equation}
p_{i}-p_{j}=\left(\sqrt{E^{2}-m_{i}^{2}}-\sqrt{E^{2}-m_{j}^{2}}\right)=\frac{\Delta m_{ij}^{2}}{2E}.
\end{equation}
We can substitute for $\Delta$ explicitly to find the oscillation phase in this model:
\begin{equation}
\frac{\Delta_{ij}}{2}L=\left(\frac{\Delta m_{ij}^{2}}{4E}+\frac{1}{2}nRe[T_{i}-T_{j}]\right)L.
\end{equation}
When the energy $E$ is very far from the resonant
energy $E_{res}^i$ for any mass state, both the new terms are negligible
relative to the standard oscillation phase, and we retrieve the usual neutrino oscillation formula.  Near resonance for a given mass component $m_k$, however, new oscillation effects become observable. The case that will be of particular interest to us is the short-baseline regime, where the standard oscillation phase can be neglected and the new terms give the dominant contribution to oscillations. 

For simplicity we will neglect CP-violation in the following discussion.  Allowing for non-zero CP violation introduces further dependencies on the CP-phases of the PMNS matrix which have not been measured,
thus introducing extra fit parameters and reducing the predictiveness of this model. These considerations would modify our results quantitatively
by O(1) numbers, though they do not change our primary conclusions.

Under these conditions, near the resonance $E_{res}^k$ and at short baseline, the new oscillation effect takes the form:
\begin{equation}
\left.P_{\alpha\beta}\right|_{L\ll L_{osc}} = 
\mathrm{sin}_{\alpha\beta}^{2}2\theta\,\mathrm{sin}^{2}\left[\frac{1}{2}nRe[T_{k}]L\right].\label{eq:OscForm}
\end{equation}
The effective angle $\mathrm{sin}^{2}2\theta_{\mu e}$ is determined entirely from
the PMNS matrix, which is assumed to be well measured by off-resonance experiments.  Resonances involving mass states lighter than the heaviest one will always be sub-dominant, due to the reduced Yukawa couplings associated with the smaller masses. Thus the dominant term for oscillation phenomenology in the majority of experimental situations arises for $k$ corresponding to the heaviest neutrino mass state.  For the normal mass ordering (NO), this is $\nu_3$, whereas for the inverted ordering (IO) it is $\nu_2$.  Thus the combination of mixing angles which determine $\mathrm{sin}^{2}2\theta_{\mu e}$ is distinct in these two cases, and given by:
\begin{equation}
\mathrm{sin}^{2}2\theta_{\mu e}^{NO}=4s_{13}^{2}s_{23}^{2}c_{13}^{2}(1-s_{13}^{2}s_{23}^{2}c_{13})=0.033
\end{equation}
\begin{align}
\mathrm{sin}^{2}2\theta_{\mu e}^{IO}=&{}4s_{12}c_{13}(c_{12}c_{23}-s_{12}s_{23}s_{13}) \times \nonumber \\
 & \left[-s_{12}c_{13}(c_{12}c_{23}-s_{12}s_{23}s_{13}\right]=0.45. 
\end{align}

Following the assumption that only one resonance is accessible in any given experiment, this model, which introduces one new particle, can be parametrized in terms of three phenomenological
parameters, $Y$, $E_{res}$ and $g_{i}$, via eq.~\ref{eq:OscProbPheno}:
\begin{widetext}
\begin{equation}
P_{\alpha\beta}=\mathrm{sin}^{2}_{\alpha\beta}2\theta\, \mathrm{sin}^{2}\left[\frac{Y}{2}\left(\frac{E-E_{res}}{\left(E-E_{res}\right)^{2}+\left(\frac{1}{2}E_{res}g_{i}^{2}\right)^{2}}+\frac{-E-E_{res}}{\left(E+E_{res}\right)^{2}+\left(\frac{1}{2}E_{res}g_{i}^{2}\right)^{2}}\right)L\right] \label{eq:OscProbPheno}
\end{equation}
\end{widetext}
with 
\begin{equation}
Y=\frac{g_{i}^{2}n}{8m_{i}} \quad\mathrm{ and}\quad E_{res}=\frac{m_{h}^{2}}{2m_{i}}.
\label{eq:YAndE}
\end{equation}

\begin{figure}
\begin{centering}
\includegraphics[width=0.49\columnwidth]{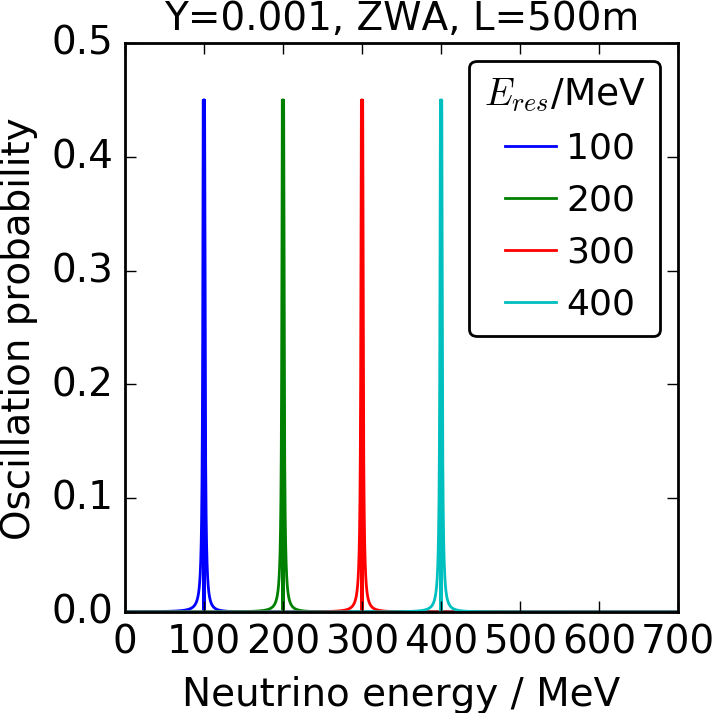}\includegraphics[width=0.49\columnwidth]{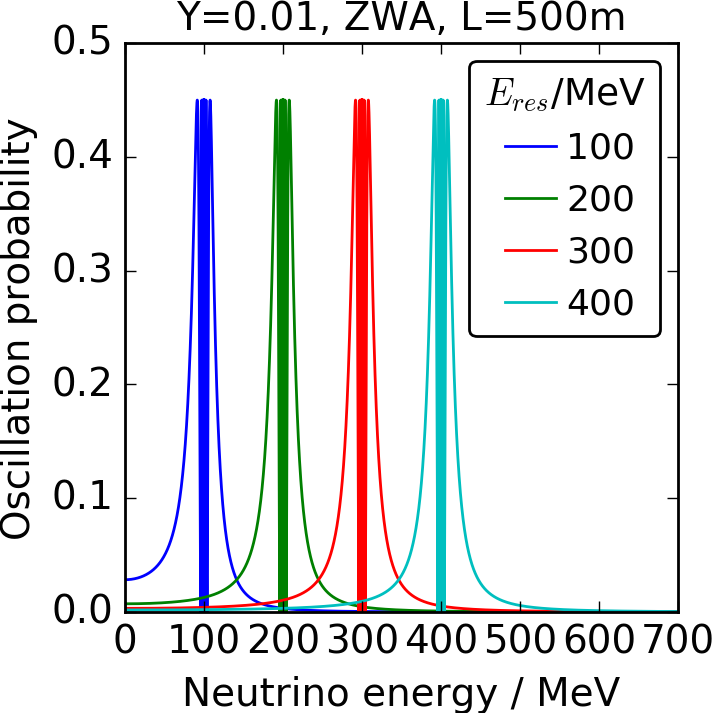}
\newline
\newline
\includegraphics[width=0.49\columnwidth]{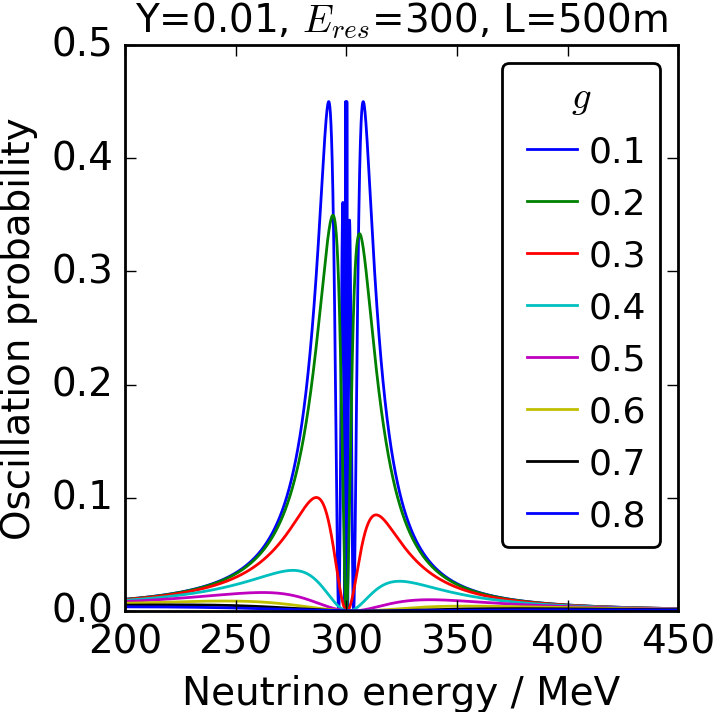}\includegraphics[width=0.49\columnwidth]{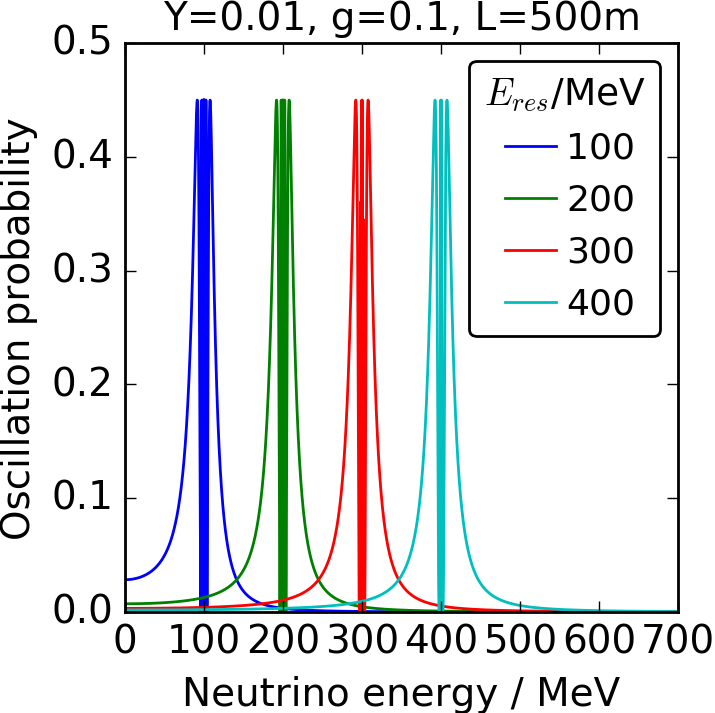}
\par\end{centering}
\caption{Example oscillation probabilities at a 500m baseline for different values of $E_{res}$, $Y$, $g$. Top: Oscillation under the Zero Width Approximation. Bottom: Oscillation for a finite width.
\label{fig:Some-example-oscillation}}
\end{figure}

We can also consider the reduced expression that applies
at low values of $g_{i}$, in practice less than $g\sim0.1$,
following from the ZWA. Under such an approximation:
\begin{equation}
P_{\alpha\beta}^{ZWA}=\mathrm{sin}_{\alpha\beta}^{2}2\theta \mathrm{sin}^{2}\left[Y\frac{E_{res}}{E^{2}-E_{res}^{2}}L\right].
\end{equation}
Some examples of the expected oscillation behavior under the ZWA are shown in Fig.~\ref{fig:Some-example-oscillation}, where a narrow peak which oscillates rapidly is clearly visible. Fig.~\ref{fig:Some-example-oscillation} also shows oscillation probabilities for some finite-width values. For narrow widths, the sharp nature of the resonance remains clear. However, for much larger values of the coupling, the
resonance becomes sufficiently wide that nontrivial oscillation structure becomes visible as the oscillation phase changes across the peak.  For the majority of the calculations described in this work, the ZWA will suffice. However, we will also briefly explore the consequences of allowing larger couplings and widths using the full model for eq.~\ref{eq:OscProbPheno}.

\section{Signatures in Short-Baseline Oscillations}

The energy-localized oscillation effect described in the previous section is particularly intriguing as a possible explanation for the short-baseline anomalies (see \cite{Abazajian:2012ys} and references therein for a full review). These anomalies suggest oscillations that are inconsistent with the understood three-neutrino paradigm. The leading phenomenological model of these observations invokes a new neutrino mass state with $\Delta m^2 \sim O(eV)^2$ and a corresponding sterile flavor state.  There is, however, an ever-growing tension between positive \cite{Aguilar-Arevalo:2013pmq,Athanassopoulos:1996jb} and null \cite{TheIceCube:2016oqi,Eitel:1998iv,Adamson:2017uda} measurements, when world data are interpreted under this model. Here we explore how the new oscillation signature derived in the previous section may explain the MiniBooNE low-energy excess \cite{Aguilar-Arevalo:2013pmq} without significant constraints from other oscillation experiments. We fit the oscillation hypothesis described above in both ZWA and finite width models, and we compare with fits with the ``industry standard'' 3+1 sterile neutrino model.

\begin{figure*}[t]
\begin{centering}
\includegraphics[width=2.00\columnwidth]{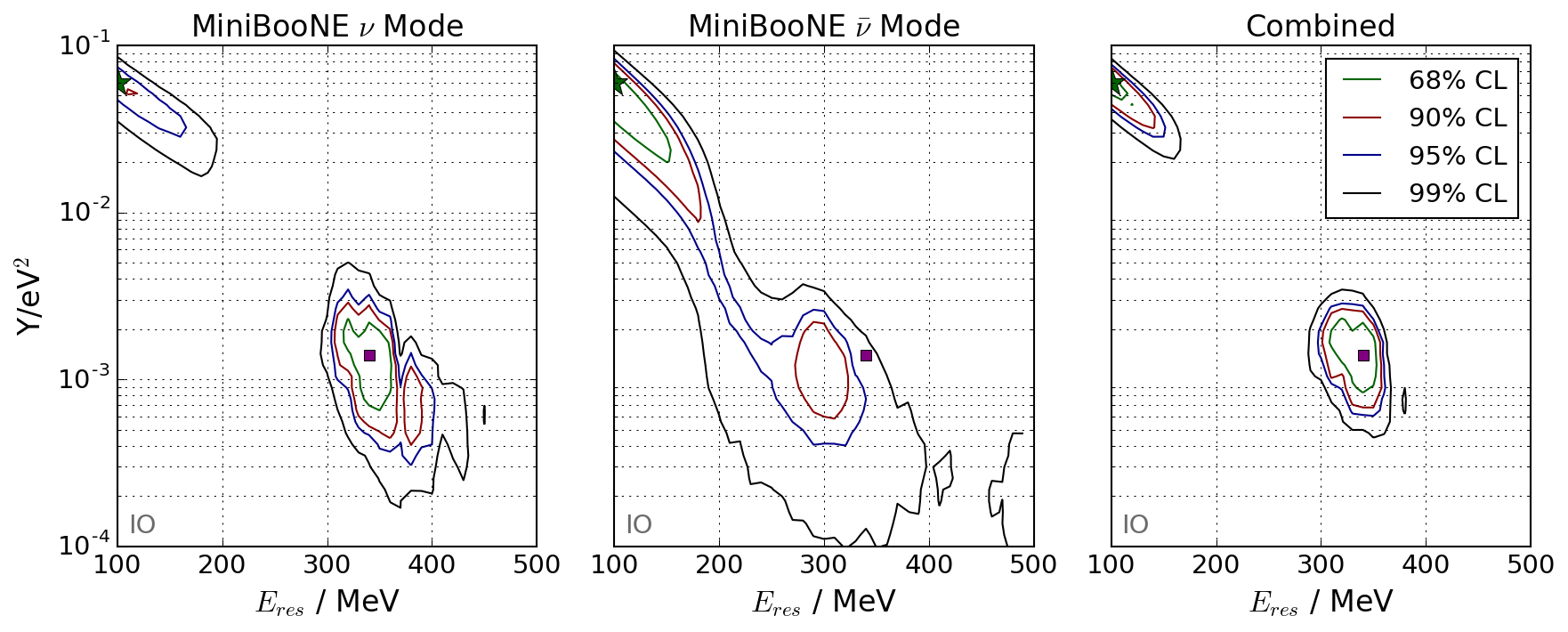}
\includegraphics[width=2.00\columnwidth]{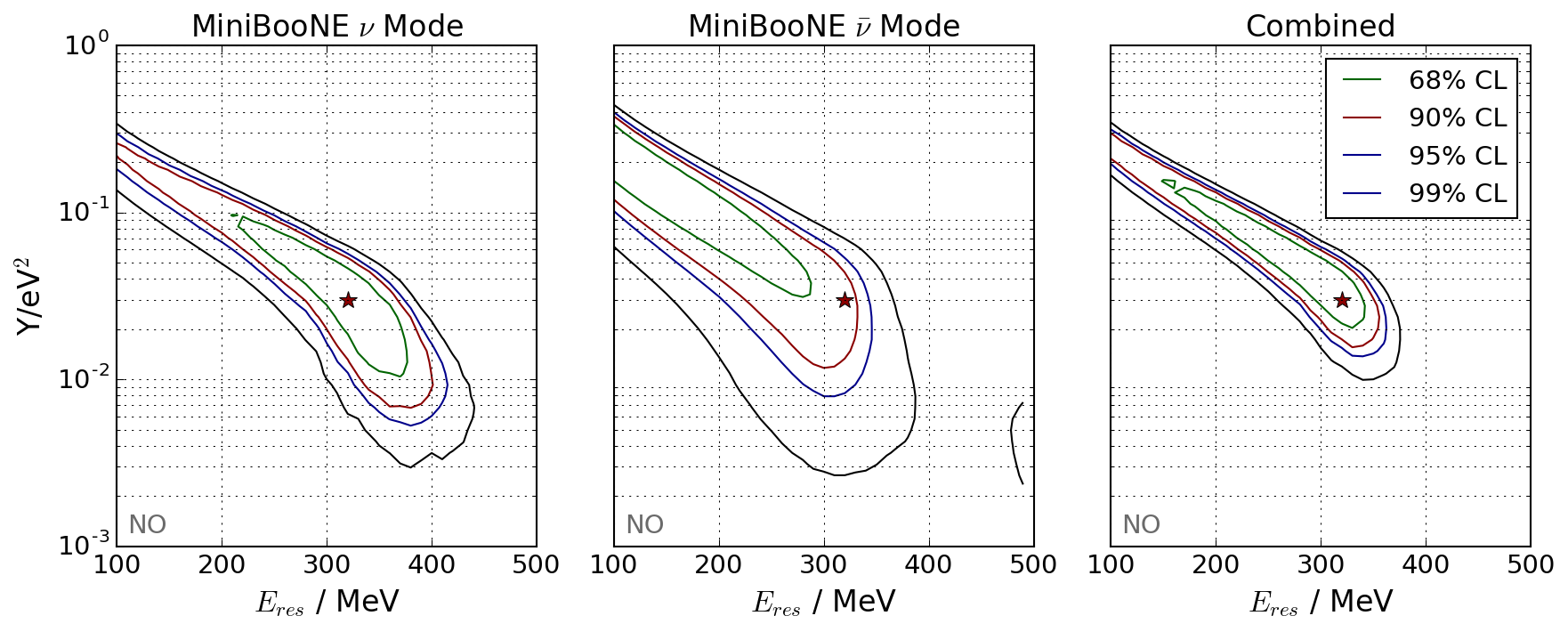}
\par\end{centering}

\caption{Allowed regions of the resonance oscillation effect in the Zero Width Approximation (ZWA) to the MiniBooNE data. Left: fits to neutrino mode only;
Center: anti-neutrino mode only; Right: combined fit. Top: inverted mass ordering; Bottom: normal mass ordering. Stars show the best fit points in each ordering, and the square shows the local minimum in the high mass island.\label{fig:AllowedRegionsInZWA}}

\end{figure*}

Our fit uses the public MiniBooNE data release accompanying \cite{Aguilar-Arevalo:2013pmq},
and employs the $\chi^{2}$ minimization recipe provided at \cite{MiniBooNEInstructions}. We employ the full MiniBooNE covariance matrix, which simultaneously fits muon- and electron- flavor samples, accounting for cross-correlations and systematic errors in both neutrino and anti-neutrino modes.
We first verified the reproduction of results under the 3+1 model, in excellent agreement with the published MiniBooNE results \cite{Aguilar-Arevalo:2013pmq}. The same fit procedure, applied to a different oscillation hypothesis, is used to identify the allowed regions using our model.  Following the example of MiniBooNE, we produce allowed regions using neutrino mode data only, using antineutrino mode data only, and using both data-sets combined.  

For the ZWA we calculate the $\chi^{2}$ value for points on a grid in ($Y$,$E_{res}$) for both the normal and inverted mass orderings and then use Wilks theorem to draw 68\%, 90\%, 95\% and 99\% confidence intervals in the ($Y$, $E_{res}$) parameter space. The allowed regions derived when fitting only MiniBooNE neutrino mode data, only anti-neutrino mode data, and both datasets together are shown in Fig. ~\ref{fig:AllowedRegionsInZWA}. The allowed regions in both modes are broadly compatible, showing the same two allowed regions overlapping strongly at 90\% CL. 

\begin{figure*}[t]

\begin{flushleft}
{\bf Neutrino Mode}
\end{flushleft}
\begin{center}
\includegraphics[width=0.67\columnwidth]{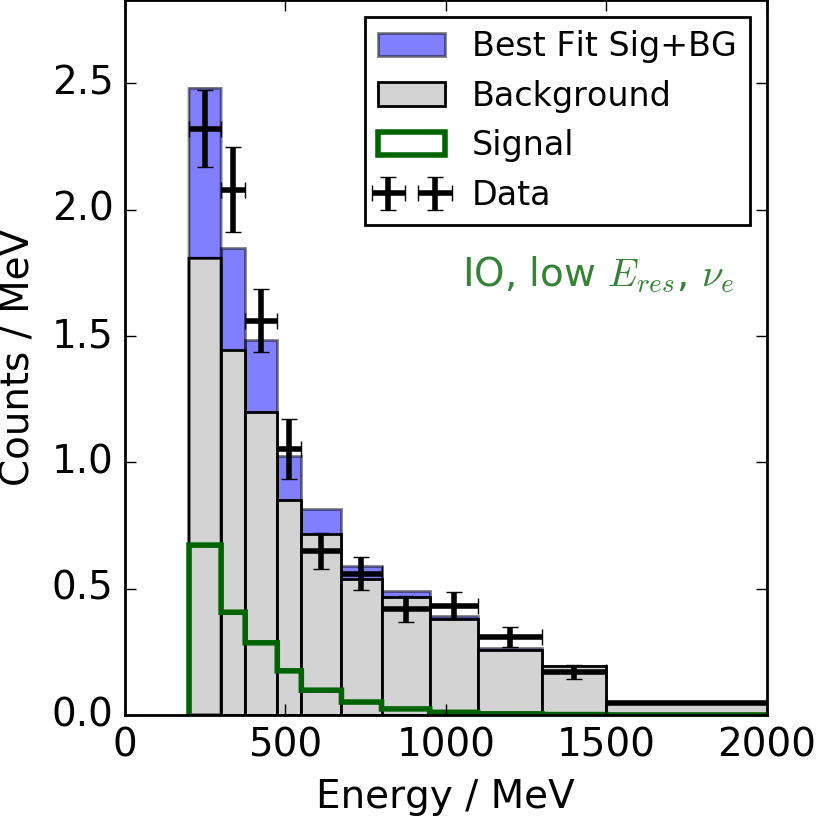}\includegraphics[width=0.67\columnwidth]{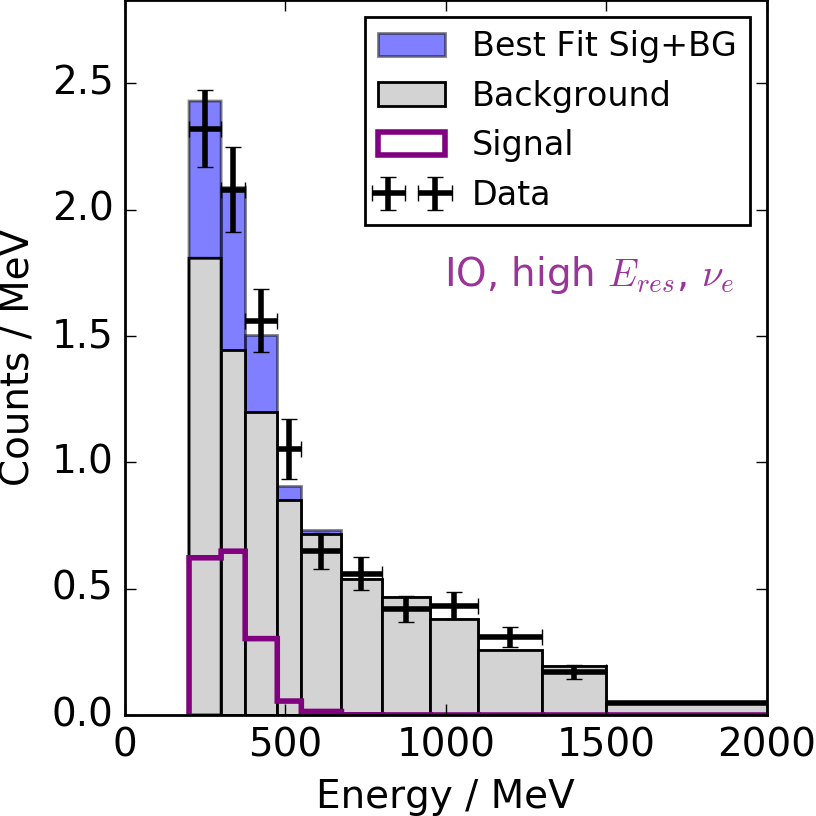}\includegraphics[width=0.67\columnwidth]{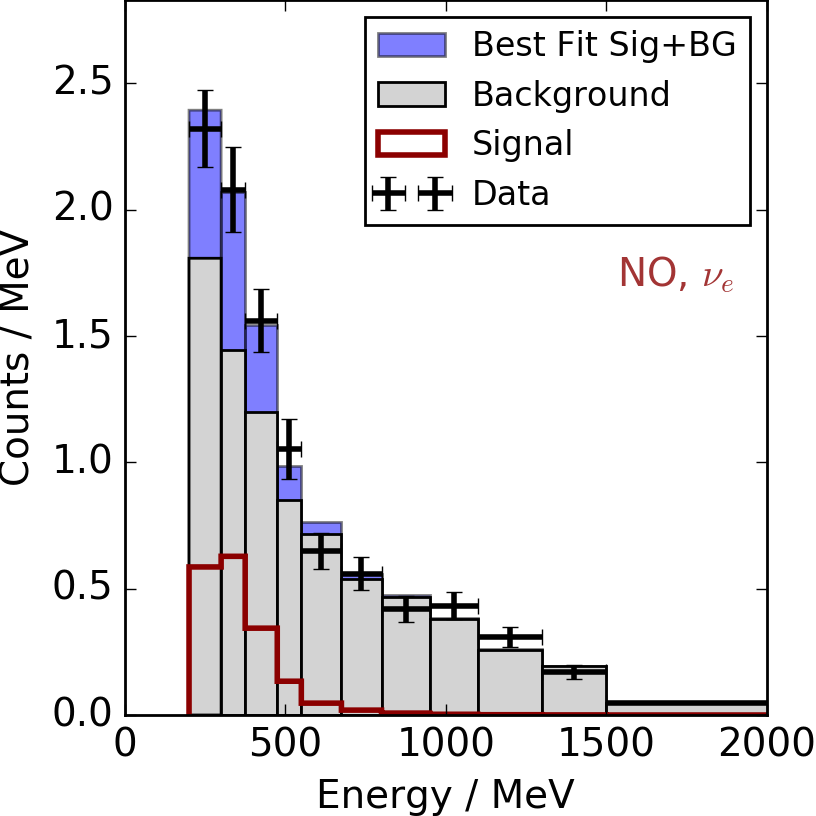}
\newline
\newline
\end{center}
\begin{flushleft}
{\bf Antineutrino Mode}
\end{flushleft}
\begin{center}
\includegraphics[width=0.67\columnwidth]{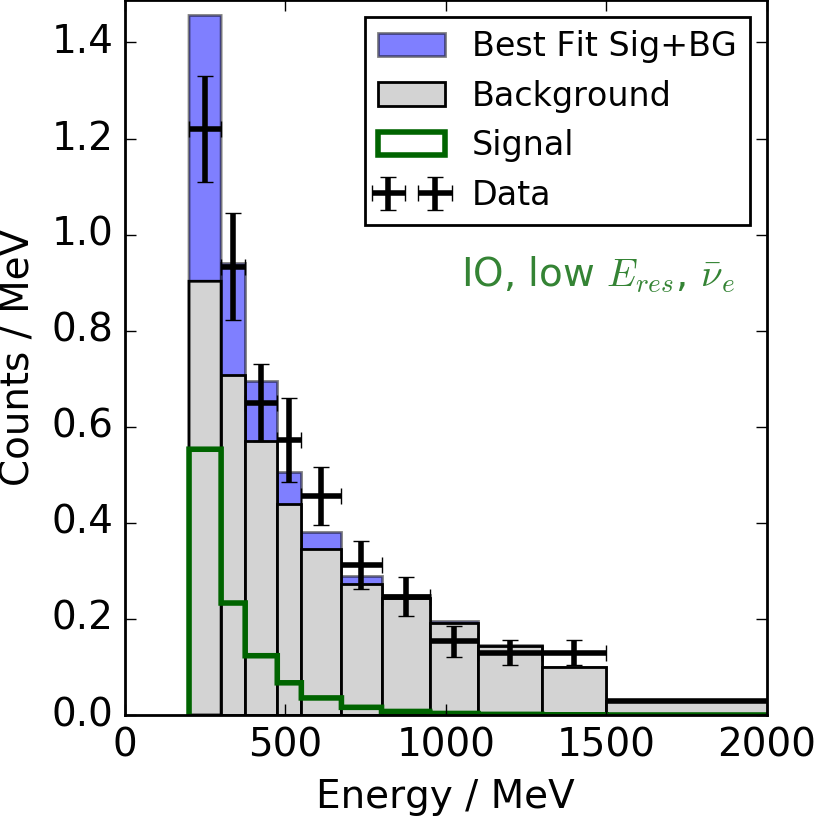}\includegraphics[width=0.67\columnwidth]{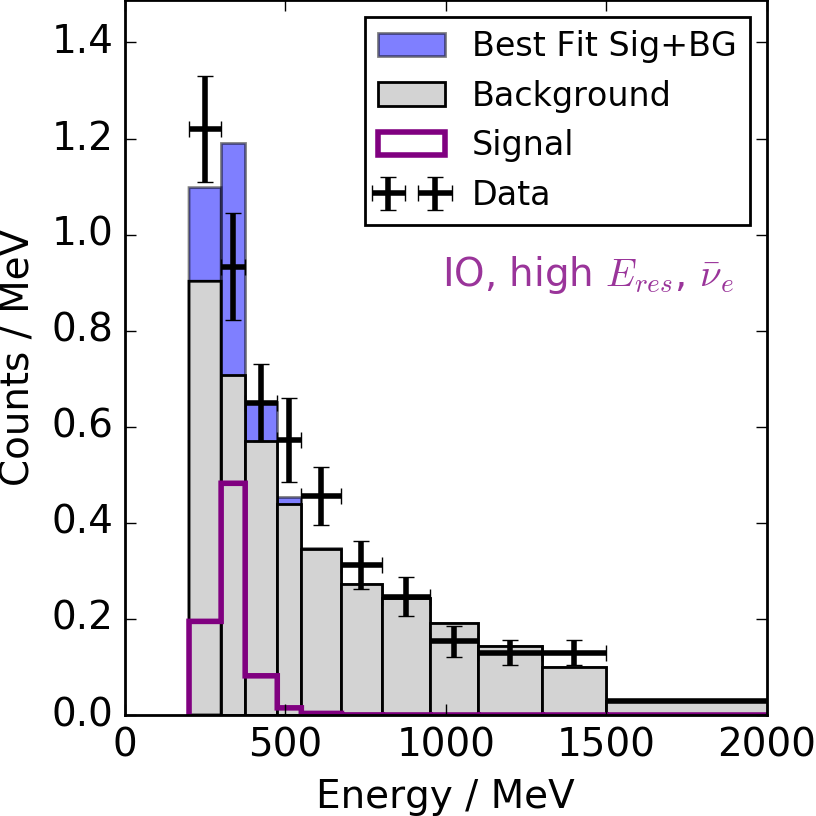}\includegraphics[width=0.67\columnwidth]{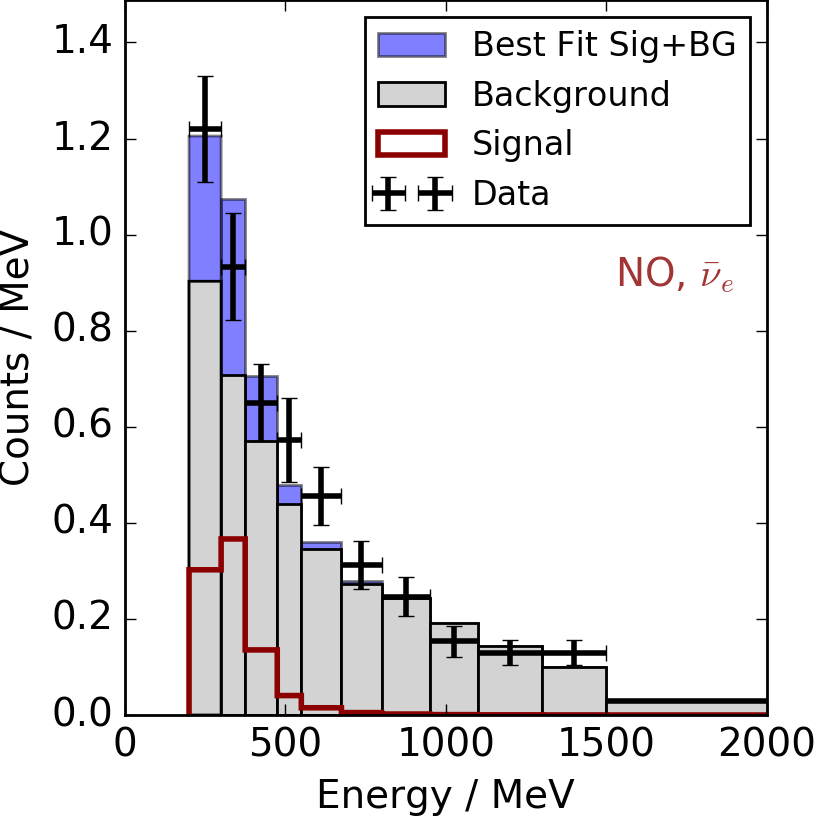}
\end{center}
\caption{Oscillation spectra in allowed regions under ZWA. Top: neutrino mode. Bottom: anti-neutrino mode. Left to right: low-mass best
fit in inverted ordering ($E_{res}=100$ MeV , Y=$6\times10^{-2}$); high-mass allowed point in inverted ordering ($E_{res}=340$ MeV, Y=$1.4\times10^{-3}$); example allowed point in normal ordering ($E_{res}=320$, Y=$3\times10^{-2}$) region.\label{fig:Best-fit-point}}
\end{figure*}

Given the inverted mass ordering, a sharp resonance around 340 MeV describes well the
data in both neutrino and anti-neutrino modes, with $\chi^{2}$ values
that indicate large p-values for the fit. A very low-mass resonance
with large amplitude also provides a good fit. For the normal ordering a wide range of resonance energies between 100 MeV and $\sim$350 give very good fits. A selection of favored oscillation hypotheses are shown
in Fig~\ref{fig:Best-fit-point}.

\begin{figure*}
\begin{centering}
\includegraphics[width=1.8\columnwidth]{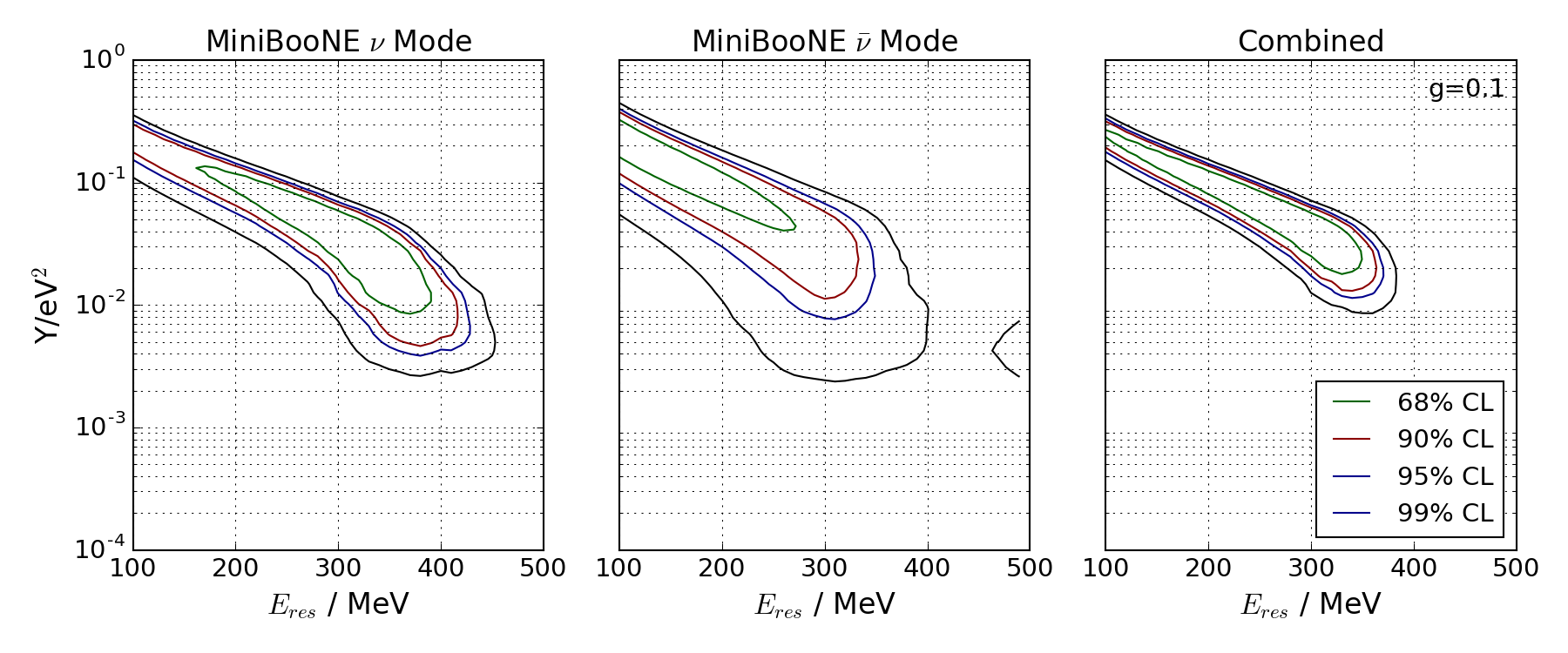}
\includegraphics[width=1.8\columnwidth]{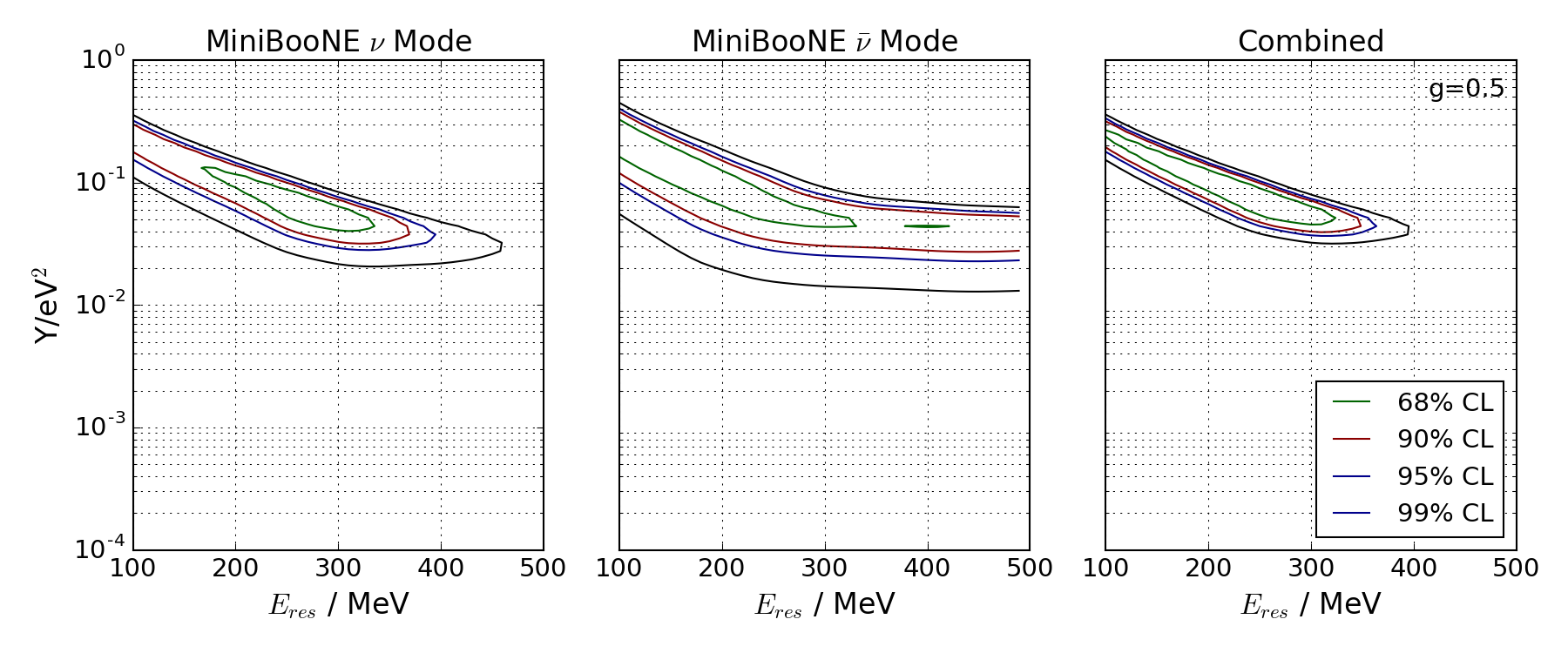}
\includegraphics[width=1.8\columnwidth]{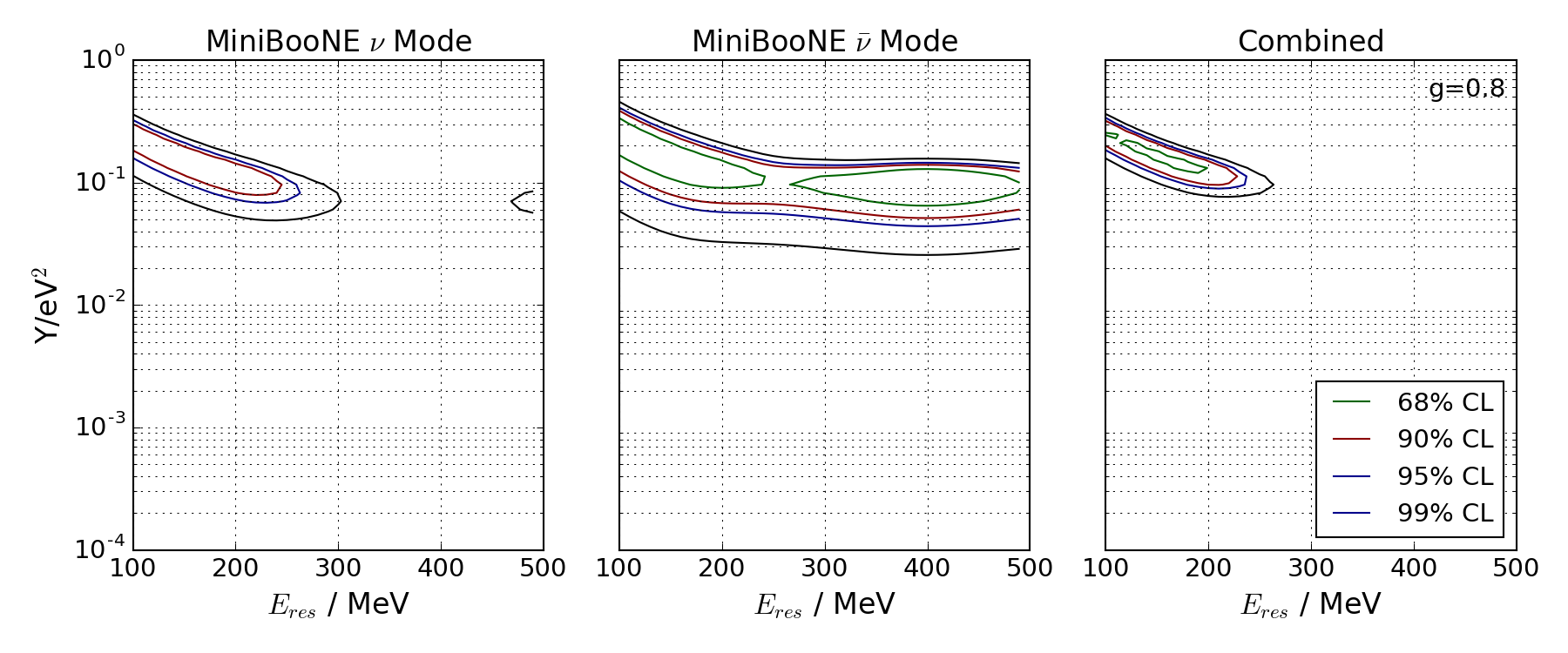}
\end{centering}

\caption{Slices of the allowed regions at fixed g given the finite width model (3~DOF) under the normal ordering. Left: fits to neutrino mode only; Center: anti-neutrino mode only; Right: combined fit. 
\label{fig:Slices-in-the}}
\end{figure*}

Relaxing the ZWA and extending the fit into the larger, 3-parameter
space allows for more flexibility to match the hypothesis at the cost
of an extra fitted degree of freedom. Slices at a few selected values of $g$ under both orderings are shown in Fig~\ref{fig:Slices-in-the}. Again, there are two allowed regions, but in this case very large values of $g$ tending to favor higher $Y$  are required, as the peak of the resonance becomes suppressed by its finite width.

Given the inverted ordering, the best fit $\chi^{2}/DOF$  (degree of freedom) in the 3-parameter fit is 24.6/35~DOF, to be
compared with 28.9/36~DOF under the ZWA.  For the normal ordering, these numbers are 24.6/35~DOF for the 3-parameter fit and 24.9/36~DOF for the ZWA. The additional fit parameter provides a modest improvement in the IO case, but not in the NO case.  However, as will be discussed in Sec.~\ref{sec:Discussion}, the large values of g required in non-ZWA scenarios are difficult to reconcile with limits on secret neutrino interactions from elsewhere~\cite{PhysRevD.90.065035}.  Since the ZWA already gives an excellent fit using one less parameter, we will consider these ZWA best fit points as the most plausible for further exploration in what follows.

\begin{table}[b!]
\begin{centering}
\begin{tabular}{|c|c|c|c|c|c|}
\hline 
\textbf{Hypothesis}& \textbf{$\chi_{\nu}^{2}$} & \textbf{$\chi_{\bar{\nu}}^{2}$} & \textbf{$\chi_{\nu}^{2}+\chi_{\bar{\nu}}^{2}$ } & $\Delta\chi_{null-bf}^{2}$ (dof)\tabularnewline
\hline 
\hline 
No Osc. & 24.2 & 23.5 & 56.0 & N/A\tabularnewline
\hline 
3+1 $\nu_s$, MB BF  & 18.8 & 12.3 & 37.4 & 18.6 (2)\tabularnewline
\hline 
3+1 $\nu_s$, global BF  & 25.5 & 19.9 & 52.6 & 3.4 (-) \tabularnewline
\hline 
$h_{\nu}$, ZWA, IO low E$_{res}$ & 11.7  & 13.3  &  30.0 &  26.0 (2)\tabularnewline
\hline 
$h_{\nu}$, ZWA, IO high E$_{res}$ & 7.9 & 20.4 & 28.9 & 27.1  (2)\tabularnewline
\hline 
$h_{\nu}$, ZWA, NO & 7.2 & 15.2 & 24.9 & 31.1  (2)\tabularnewline
\hline 
$h_{\nu}$, IO full & 7.8 & 14.5 & 24.6 & 31.4 (3)\tabularnewline
\hline 
$h_{\nu}$, NO full & 7.6 & 14.6 & 24.6 & 31.4 (3)\tabularnewline
\hline 
\end{tabular}
\par\end{centering}

\caption{$\chi^{2}$ values for null hypothesis, the best fit
in the 3+1 model using MiniBooNE and the global fit value from \cite{Collin:2016rao}, and several of the fit points in the neutrino-Higgs model. In the latter case we consider both the absolute best fit (low mass),
the best fit point in the high mass island for the IO, an allowed value for the NO, and the best fit in the
finite width case. These fits use 19 bins in each of neutrino and antineutrino modes (11 $\nu_e$ and 8 $\nu_\mu$) and data from \cite{Aguilar-Arevalo:2013pmq}. \label{tab:-values-and}}
\end{table}

In both ZWA and finite-width models, this new light Higgs hypothesis gives an improvement over the 3+1 model for fitting
the MiniBooNE excess. The 3+1 model taken at world-best fit from \cite{Collin:2016aqd} is only very slightly favored by MiniBooNE data relative to no oscillations, by 3.4 points in $\Delta\chi^2$.  The fit to MiniBooNE data only, allowing for parameters disfavored by world data, provides a $\Delta \chi^2$ improvement of 18.6 with 2 DOF.   The much larger $\Delta\chi^2$ values for all versions of the neutrinophilic Higgs model (with 2~DOF under the ZWA or 3~DOF with the full finite-width model), of 26 $\leq \Delta \chi^2$ $\leq$ 31.4, show that it is a significantly better fit to the MiniBooNE data.  

Table \ref{tab:-values-and} gives a comparison
at the best-fit point fitted using combined neutrino and anti-neutrino
datasets, evaluated in neutrino mode, in anti-neutrino mode, and in both together between various possible models of the excess and the null hypothesis. Our model appears significantly preferred over both the null hypothesis and the sterile neutrino explanation, both at its best-fit point from MiniBooNE only, and at the world-data best fit.

As has been discussed by other authors \cite{Maltoni:2003cu}, one should exercise caution in the interpretation of overall $\chi^2$ as a quantitative goodness-of-fit for each model, since this fit considers both muon- and electron-flavor data simultaneously, only a subset of which is expected to exhibit signal.  Fluctuations in the $\chi^2$ at the expected level from non-signal bins can hide even large anomalies in localized regions. Rather, the key metric for comparing models is $\Delta\chi^2$, which if significantly larger than the number of degrees of freedom in the model, indicates an improvement of compatibility with data over the null hypothesis.  Both 3+1 and light Higgs models, when unconstrained by other oscillation data, provide significant improvements. Furthermore, the light Higgs fits substantially out-perform the 3+1 model fits in all favored regions.

\begin{figure*}[t]
\begin{center}

\includegraphics[width=0.704\columnwidth]{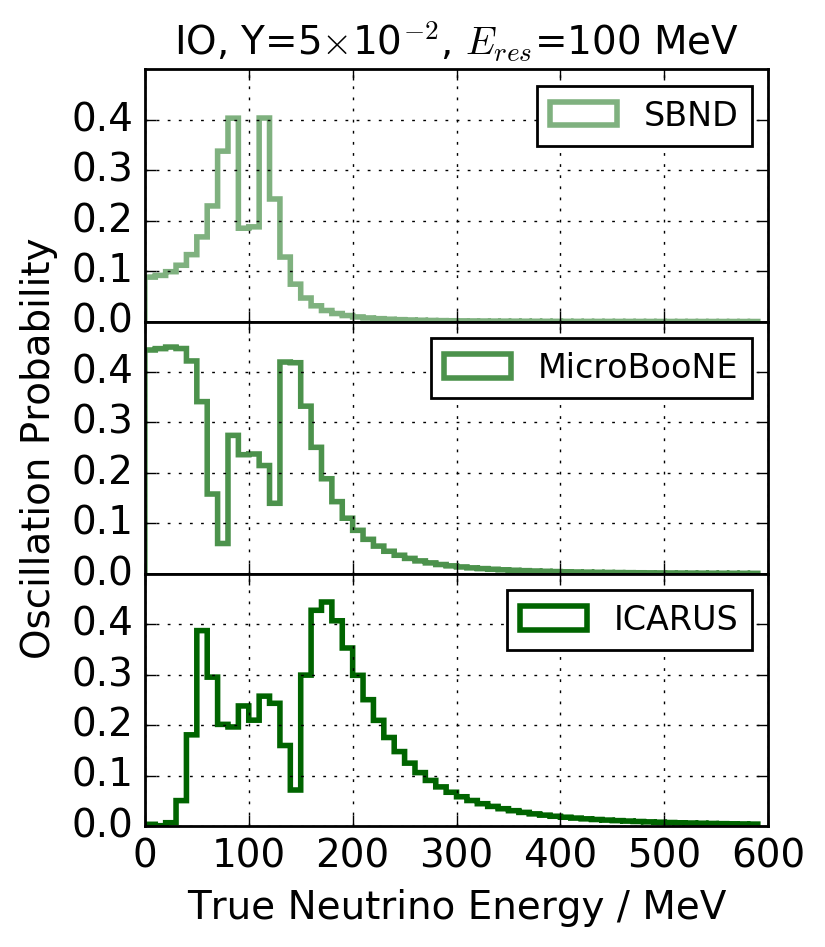}\includegraphics[width=0.673\columnwidth]{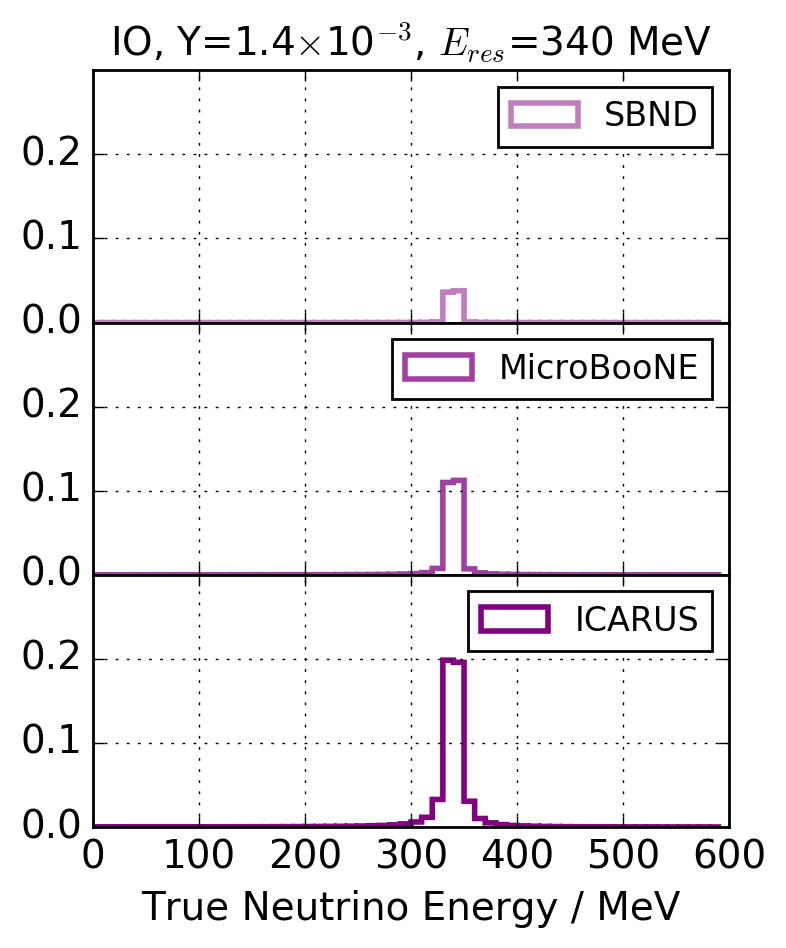}\includegraphics[width=0.673\columnwidth]{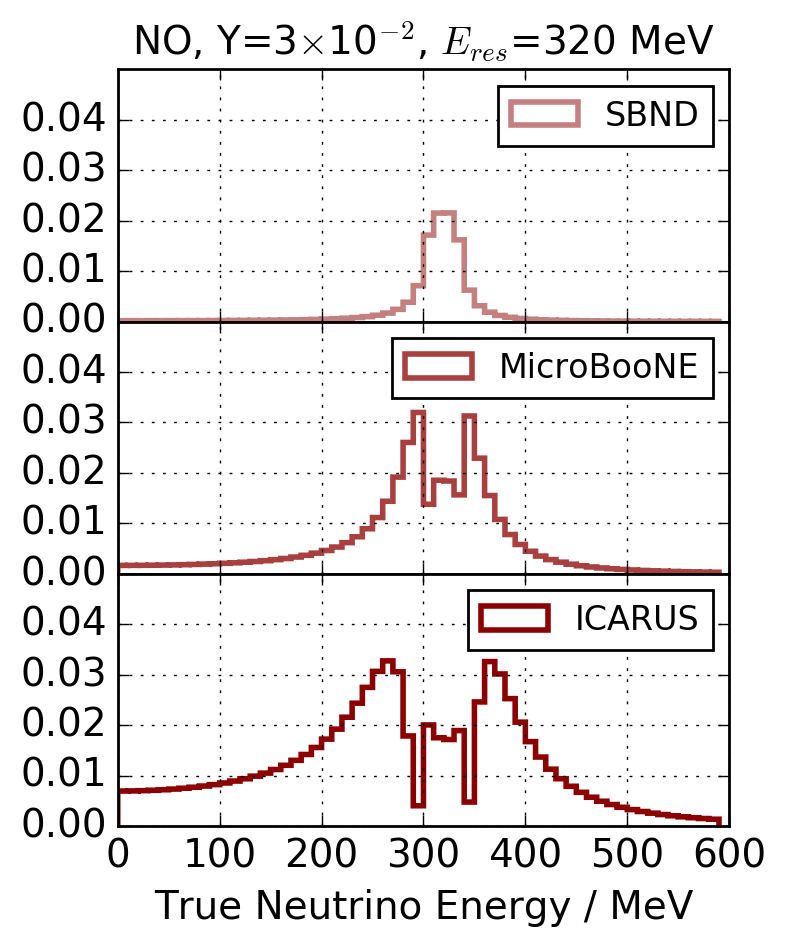}

\end{center}
\caption{Predictions for the oscillation probability for the SBN program for the favored high and low energy resonance solutions in IO (left and center) and an example allowed point in the NO (right).\label{fig:SBNPredictions}}
\end{figure*}

\section{Implication for Future Short-Baseline Neutrino Experiments}

The Short-Baseline Neutrino (SBN) program at Fermilab \cite{SBN} has been proposed to directly address the short-baseline oscillation anomalies through the combination of three liquid argon time projection chambers located along the Booster Neutrino Beam. In addition to the oscillation searches from $\nu_e$ appearance and $\nu_{\mu}$ disappearance, the SBN also offers sensitivity to beyond the standard model searches such as heavy sterile neutrino decay searches \cite{PhysRevLett.103.241802, PhysRevD.85.051702, NuBounds} and sub-GeV dark matter searches running in beam dump mode \cite{PhysRevD.80.095024, PhysRevD.84.075020, PhysRevD.86.035022, PhysRevLett.118.221803}. These searches are enhanced by the presence of the three detectors because of the baseline dependence of oscillations and the attendant systematic uncertainty reduction afforded by a multi-detector program. The oscillation effect described in this paper would be directly observable by the SBN detectors, and clearly distinguishable from the 3+N sterile neutrino scenario. Fig.~\ref{fig:SBNPredictions} shows the prediction for the oscillation effect described here at each SBN detector baseline for each of the three favored regions in Figure~\ref{fig:AllowedRegionsInZWA}.

The extrapolated oscillations corresponding to each solution derived from MiniBooNE data show clear and distinctive features in the three SBN detectors. In the case of the low energy IO resonance solution, the oscillation spectrum in each of the detectors has a very different shape. The difference in this spectrum shape provides a powerful handle to identify and classify the effect described here. In the case of the high energy resonance IO solution, the peak at 340 MeV in all detectors would be a clear signature of this effect. The NO solution has a smaller amplitude but a non-trivial energy structure at 300~MeV.  The energy regime and structure of these effects is ideally suited to exploration at the SBN program, since these detectors are expected to have high efficiency and excellent calorimetric resolution in this energy range.  Moreover, the  three-detectors configuration allows for the oscillation probability to be mapped as a function of baseline, allowing for detailed exploration of exotic oscillation scenarios such as this one. The combination of three SBN detectors presents the opportunity for a conclusive test of the phenomena presented here, as well as other baseline dependent phenomena arising from exotic neutrino oscillation scenarios.

\section{Discussion \label{sec:Discussion}}

We have shown that the MiniBooNE low energy anomalies can be explained using model with a new, light boson that couples to neutrinos. We demonstrated that the new interaction introduces a resonance in the neutrino oscillation probability due to 
the forward scatter of neutrinos from an over-dense relic neutrino background. The allowed regions in our fits suggest that resonances
around $E_{res}\sim100$ MeV and $E_{res}\sim340$ MeV provide the
optimal fit points in the case of the inverted neutrino mass ordering, and for a wide range of masses up to 350~MeV given the normal neutrino mass ordering. This resonance energy is related to the mass scale
of the new boson by $E_{res}^i=m_{h}^{2}/2m_{i}$

If the MiniBooNE anomaly is to be explained via this effect, we expect
that it must be the heaviest neutrino mass state which is on resonance,
in order to avoid conflict with non-detection of larger and higher-energy resonance peaks in other experiments. The
mass of this neutrino is constrained from oscillations and
cosmology, and the resonance energy leads us to conclude that the mass of this new Higgs boson would be either 6-9 keV in the case of the higher energy class of solutions, or 3-5 keV for the lower energy class. 

Of the existing oscillation searches, only the T2K experiment has significant sensitivity in this energy range \cite{T2K2017}. Because the maximum oscillation amplitude under the normal mass ordering is small, all hypotheses in the normal ordering scenario are outside the sensitivity of T2K.  For the inverted ordering solutions, the effective mixing angle is larger and exploration with T2K is possible.  Sensitivity depends on reconstruction efficiency, energy resolution and uncertainty budget in the energy range of interest.  These pieces of information are not presently available at the level of detail required to make a careful numerical estimate.  However, evaluation of eq.~\ref{eq:OscForm} at 295km, and comparison to T2K published data \cite{T2K2017} strongly suggests that the solutions with large resonance energy, which have smaller $Y$, will not be excluded within the stated uncertainties.  The low mass solutions with larger $Y$ produce a wider resonance at long baselines, which is likely to be at least partially constrained by T2K.  Notably, the MINOS \cite{Adamson:2013ue} and NO$\nu$A \cite{Adamson:2017gxd} experiments do not have high enough statistics at low enough energies to address these models with their published analyses, though dedicated searches with cuts optimized for low energy $\nu_e$ acceptance may be possible. 

For the low energy inverted ordering solution, the other prominent experiment in the appropriate energy range is LSND \cite{Athanassopoulos:1996jb}, which also observes an excess in $\bar{\nu}_e$ appearance. That excess could be explained by this mechanism, though as with any short-baseline oscillation hypothesis, some fine tuning to maintain consistency with null results from KARMEN \cite{Eitel:1998iv} would likely be required.  A full evaluation of the constraints imposed by world oscillation data would require comprehensive global fits, as have been performed for the 3+1 sterile neutrino hypothesis \cite{Collin:2016rao,Kopp:2013vaa,Gariazzo:2015rra}.  This activity is outside the scope of the present work.

Fitted values of $Y$ of $10^{-3}$ and larger are allowed at
90\% CL.  The phenomenological parameter $Y$ is related to the coupling strength, neutrino mass and local neutrino over-density via eq.~\ref{eq:YAndE}.  Taking the heaviest neutrino mass at its smallest allowed value of
0.05 eV, a relic density $\geq4\times10^{-4}$ eV$^{3}$ is implied. This exceeds the standard expected global relic density by a factor
of order $10^{8}$. This is a large value, but is smaller than densities that have been discussed in other literature \cite{Kaboth:2010kf,Hwang:2005dq}, and remains outside of direct experimental limits. These levels of over-density should be observable by future proposed experiments \cite{Betts:2013uya}. Constraints on ``secret'' neutrino interactions (compiled in~\cite{PhysRevD.90.065035})  further bound the range of values for $n$ that are consistent with our fitted parameters.  These constraints fall into two broad classes: direct limits on $g_i$  from particle decay widths \cite{Bilenky:1999dn,Lessa:2007up}, and limits from absorption or down-scattering of neutrinos travelling over astronomical baselines proportionally to $\eta=g^4_in$ \cite{Kolb:1987qy,PhysRevD.90.065035}.  The refractive mechanism described here, on the other hand, depends on $g^2_in=\eta/g^2$.  Hence both of the above are evaded with a suitably large local over-density and small $g_i$.  

There are also possible implications of a new light scalar for cosmology \cite{PhysRevD.52.1764, NuBounds, Hannestad:2005ex}.  Naively this particle contributes $\Delta N_{eff}$ = 4/7 of a neutrino-equivalent degree of freedom in the early Universe.  This is consistent with measurements of big bang nucleosynthesis at the 1.5$\sigma$ level~\cite{Patrignani:2016xqp,Peimbert:2016bdg,Aver:2015iza,Izotov:2014fga}. The Planck~\cite{Ade:2015xua} determination of $N_{eff}$=3.04$\pm$0.18 from the CMB,  taken at face value,  would appear to be a stronger constraint.  However, this measurement is strongly correlated with the measured Hubble constant, which is in significant tension with direct determinations \cite{Efstathiou:2013via,Riess:1998cb,Freedman:2012ny}. Fixing the value of $H_0$ to that measured elsewhere would substantially weaken the constraint on $N_{eff}$, pulling it to higher values, and thus drawing the robustness of this constraint into question.

Although preliminary indications suggest that the constraints from cosmology are not overly severe, the extent to which this particle is in equilibrium, and the impact of its possible late-era phase transition, may complicate the simple interpretation of these constraints given above. We thus defer a full account of cosmological implications of this model to future work.

\section{Conclusions}

In this paper we have presented the phenomenological implications of a light, neutrinophilic Higgs boson, motivated as a mechanism to generate neutrino mass. A resonance effect from a neutrino beam interacting with an over-dense relic neutrino background is explored that can provide an alternate explanation to short-baseline oscillation anomalies. This model provides a good fit to the MiniBooNE low-energy excess and illustrates an interesting class of new oscillation phenomena. This may provide a promising direction for new physics searches at the Fermilab Short-Baseline Neutrino program and beyond.

\section*{Acknowledgements}

We thank Carlos Alberto Arg\"{u}elles for his insightful comments throughout; Boris Kayser for illuminating discussions about neutrino-Higgs couplings; and Fernanda Psihas for consultations about long-baseline experiments and for a thorough proof-read of this manuscript. Thanks also to Joachim Kopp and Pedro Machado for their comments on the pre-print, which were accommodated into this work before final submission.  The majority of this work was undertaken at the first and second ``Rencontres de Taos'' workshops in 2016 and 2017, and we thank the organizers and FMT for their efforts. A.M.S. is supported by a Royal Society University Research Fellowship.


\bibliographystyle{apsrev4-1}
\bibliography{main}

\end{document}